\documentclass{aa}
\usepackage{natbib, my_def, graphicx} 
\bibstyle{aa}     
\def\udier{\"u}

\def \etal   {{~et~al.~}}
\begin{document}
\title{VLBA polarimetric observations of the CSS quasar 3C\,147}
 
\author{ A. Rossetti \inst{1} \and
         F. Mantovani  \inst{1} \and
         D. Dallacasa \inst{1,2} \and  
	 W. Junor \inst{3}   \and
	 C.J. Salter \inst{4}     \and
	 D.J. Saikia \inst{5}
         }

\offprints{A. Rossetti,\\
  \email{rossetti@ira.inaf.it}}

\institute{Istituto di Radioastronomia -- INAF, via Gobetti 101,
 I--40129, Bologna, Italy 
 \and Dipartimento di Astronomia, Universit\`a degli Studi, via
 Ranzani 1, I--40127 Bologna, Italy  
 \and Los Alamos National Laboratory, Los Alamos, NM 87545, USA
 \and Arecibo Observatory, HC3 Box 53995, Arecibo, Puerto Rico 00612
 \and National Centre for Astrophysics, TIFR, Post Bag 3,
 Ganeshkhind, Pune 411 007, India
}

\date{Received \today; accepted ???}

\abstract
{}
{We report new VLBA polarimetric observations of the compact
steep-spectrum (CSS) quasar 3C\,147 (B$0538+498$) at 5 and 8.4~GHz. }
{By using multifrequency VLBA observations, we 
derived milliarcsecond-resolution images of the total intensity,
polarisation, and rotation measure distributions, by combining our new
observations with archival data.}
{The source shows a one-sided structure, with a compact region, and a component extending about
200 mas to the south-west. The compact region is resolved into two main
components with polarised emission, a complex rotation measure
distribution, and a magnetic field dominated by components
perpendicular to the source axis.}
{By considering all the available data, we examine the possible
  location of the core component, and discuss two possible
  interpretations of the observed structure of this source: core-jet
  and lobe-hot spot. Further observations to unambiguously determine
  the location of the core would help distinguish between the two
  possibilities discussed here.}


\maketitle 
\section{Introduction} 
\label{sec:intro}

Compact steep-spectrum (CSS) radio sources are scaled-down
versions of large-sized double sources with linear sizes $\leq
20\,h^{-1}$\,kpc \footnote {$q_0=0.5$ and $H_0=100\,h\,{\rm km}\,{\rm
  s}^{-1}\,{\rm Mpc}^{-1}$} and steep high-frequency radio
spectra ($\alpha >0.5$ \footnote{we assume ${\rm S}_{\nu}=\nu^{-\alpha}$}).
It is now generally believed that most CSS sources are {\it young},
of ages $<10^{3-5}$\,yr, whose radio lobes have had 
insufficient time to grow to kilo-parsec scales ~\citep{Fanti95}.
Subgalactic in size, they reside deep inside their host galaxies
and are largely confined to the Narrow Line Region (NLR) with its
relatively large column density of ionised plasma.
This magnetised thermal plasma has different indices of refraction for the two circular
polarised modes of the synchrotron emission of these sources. This
rotates the orientation of the linearly polarised
components of the transmitted radiation. The amount of the rotation is
given by $\Delta\chi=812\times\lambda^2\,\int{n_{\rm e} B_{||}
d\ell}$, where $\lambda$ is the wavelength in cm, $n_{\rm e}$ is the
electron density  of the medium in cm$^{-3}$, $B_{||}$ is the component of the 
magnetic field along the line of sight in $\mu$G,  and $\ell$ is the
geometrical depth of the medium along the line of sight in kpc. 

Since the rotation is proportional to $\lambda^2$, it can be
determined from observations at a number of wavelengths. The Faraday
{\it Rotation Measure}, {\it RM}, is the amount of rotation expressed
in rad$\,$m$^{-2}$. Even for moderate magnetic field intensity for the
foreground screen,
significant Faraday rotation effects external to the radio sources are
to be expected, which can completely depolarise sources at long
wavelengths. 
The comparison of polarisation images over a range of wavelengths,
preferably with similar resolutions, is then an
important diagnostic of the physical conditions within and around CSS
radio sources. 
CSS radio galaxies show very low, or even no, linear polarisation at
centimetre wavelengths. CSS quasars are instead found to have
polarisation percentages as high as 10\% above 1~GHz \citep{Saikia85,
  Saikia87, Fanti04, Rossetti08}.

A minority of CSSs exhibit complex or highly-asymmetric structures.
Observations indicate that intrinsic distortions are caused by
interactions with a dense, inhomogeneous gaseous environment
~\citep{Mantovani94}.  
This view is supported by the most distorted and complex structures being found in
jet-dominated objects with very weak cores ~\citep{Mantovani02}. 
Furthermore, an increased asymmetry in terms of intensity, arm ratio, spectral index, and
polarisation with decreasing source linear size suggests that they
are expanding through the dense inhomogeneous interstellar medium
(ISM) of their host galaxies ~\citep{Sanghera95, Saikia01, Saikia03, Rossetti06}. 

Sub-arcsec polarimetry provided evidence of
the interaction of components of CSS sources with dense clouds of gas
~\citep{Nan99}.  
According to ~\citet{Junor99a}, the CSS source \object{3C\,147} has the
expected signatures of a jet colliding with a cloud of gas on the
southwestern side of the galaxy. 
Optical observations also show evidence of such interactions. 
\citet{GW94} found that CSS sources have relatively strong, high
equivalent width, high excitation line emission, with broad,
structured [OIII]\,$\lambda\,5007$ profiles, which they interpret
as evidence of strong interaction between the jet and the ISM.  

The number of CSSs for which detailed radio polarisation information is
available remains small. To improve the statistics,
we are conducting a series of observations to image examples 
with moderate degrees of polarised
emission and clear signatures of interaction with
their environment, namely fractional polarisations that decrease with
increasing wavelength, and $RM>450$\,rad\,m$^{-2}$ in the source rest frame. 
Results for the first two CSS quasars observed in
 this ongoing program, \object{B$0548+165$} and \object{B$1524-136$},
 are available in  ~\citet{Mantovani02}.  

We present results for \object{3C\,147} (\object{B$0538+498$}), imaged with
milliarcsecond resolution via full-Stokes VLBA observations.
This CSS radio source is associated with a QSO of
$m_{v}=16.9$ and $z=0.545$ \citep{Spinrad85}. 
It has a total angular extent of $0\farcs 6$, corresponding to a
projected linear size of $\approx 2.2\,h^{-1}\,{\rm kpc}$. 
\object{3C\,147} shows a very large integrated {\it RM} of $\approx
-1500\,{\rm rad}\, {\rm m}^{-2}$ in the observer's frame
\citep{Inoue95}, implying that the radio source is surrounded by 
a dense ionised medium.
On the sub-arcsecond scale \object{3C\,147} exhibits a
very asymmetric double-lobed radio structure with a strong,
steep-spectrum, central component and asymmetric components of even
steeper spectrum situated to the southwest and north \citep{A-G95,
Ludke98, Junor99a}.  
An HST optical study of the emission-line region of CSS radio sources
~\citep{Axon00} shows that the emission-line
morphology ([OIII]\, $\lambda\,5007$) of \object{3C\,147} is 
dominated by an unresolved central component, embedded in a more
diffuse region, which extends over $\approx 1\farcs2$ and is well
aligned with the double radio structure observed on the sub-arcsecond scale. 

 \citet{Junor99a} imaged the polarised emission of \object{3C\,147} using
 A-array VLA observations at X- and U-bands. They measured significant
differential Faraday rotation between the northern and the 
main (central+south-western) components situated on opposite sides of
the quasar, and a strong east-west gradient in the polarisation
percentage of the main component, whose emission depolarises strongly towards
lower frequencies.  They interpreted these results in terms of an
asymmetric environment with large amounts of thermal gas on the western
side.

Very Long Baseline Interferometry (VLBI) observations have been analysed to indicate that
the central component contains a compact core with a jet emerging
from it toward the southwest \citep{Alef90, Nan00}. Part of this jet
yields evidence of mild superluminal motion: \citet{Alef90} reported 
$\approx 1.3\,c$, and \citet{Nan00} inferred $\approx 1.5\,c$.  

In this paper, we present multi-frequency VLBA (plus a single VLA antenna)
polarisation observations at 5 and 8.4~GHz. 
In Sect. \ref{sec:observation} we
summarise the observations and data processing. Section
\ref{sec:results} describes 
new information obtained about the morphological and polarisation properties of
\object{3C\,147}. Discussion and conclusions are presented in Sects.
\ref{sec:discussion}  and \ref{sec:conclusions}.  

\section{Observations and data reduction} 
\label{sec:observation}

\subsection{5~GHz and 8.4~GHz data}
\label{subsec:myobservation}

Polarimetric observations of \object{3C\,147} were carried out in May 2001
at 5 and 8.4~GHz for about 12 hr with the VLBA plus one VLA
antenna. This array has the highest quality observing capability for detecting the
extended, complex, milliarcsecond structure of this source.
The data were recorded in both right- and left-circular polarisation
in four 8-MHz bands at each frequency. 
At 5~GHz, the observations were spread across the available 
bandwidth of $\approx 500$~MHz, allowing us to obtain four truly simultaneous
and independent polarisation images and study the rotation
measure distribution. To increase the sensitivity to the
polarised emission, at 8.4~GHz we decided to use contiguous
IFs. The resulting Intermediate
  Frequencies (IFs) are 4619\,MHz, 4657\,MHz, 4854\,MHz, and
  5094\,MHz at C band, and 8405\,MHz, 8413\,MHz, 8421\,MHz, and
  8429\,MHz at X band. 
 The data were correlated with the NRAO VLBA processor at Socorro and
calibrated, imaged, and analysed using the AIPS package.

After the application of system temperature and antenna gain
information, the amplitudes were checked by means of the data on the
source  \object{DA\,193}, which was used as flux density calibrator, fringe
  finder, and polarisation calibrator. 
All the gain corrections were found to be within the $3\%$, which can
be conservative assumed as absolute calibration uncertainty. Standard
procedures were applied for the fringe 
fitting of both total intensity and polarised signal. 

During the data reduction process, the data at each of the four
IFs within the C-band were processed separately. The polarised
source \object{DA\,193} was used to determine the instrumental
polarisation (``D-term'') using the AIPS task PCAL. The solution shows
that the instrumental polarisation was typically of the order of
$1\%$.
After the  D-term calibration, there remains an arbitrary offset in
the polarisation position angle, $\chi$, which can be determined by
using integrated measurements of the
source \object{DA\,193} from the NRAO polarisation database,
assuming that the 
source emission is confined to within its milliarcsecond
structure. Since \object{DA\,193} exhibits
time variability and there were no  
NRAO measurements close to our observing epoch (2001 May 6), we
had to extrapolate the polarisation of \object{DA\,193} using a
  linear interpolation of all the measurements
available for this source from 2000 to 2002 at the VLA/VLBA polarisation calibration
page \footnote {http://www.vla.nrao.edu/astro/calib/polar/}. 
To remove any offsets between the four IFs at
C-band, we first 
rotated the phases of the polarisation {\it uv} data for all
sources so that the $\chi$ values for \object{DA\,193} at each IF were aligned
with the value of IF1 (4619\,MHz). This procedure assumed that the Faraday rotation 
of \object{DA\,193} across the observing bandwidth was negligible, a
reasonable assumption given the low integrated {\it RM} of
this source ($\approx 107 \,{\rm rad}\,{\rm m}^{-2}$). A systematic error of
$\approx 5\degr$ was introduced over the whole bandwidth. As a result,
there is an uncertainty in the {\it RM} within the C-band. 
As an overall check of the reliability of our calibration process, we compared
the results obtained for the source \object{$0927+390$} with those found in 
 the NRAO data archive.
The polarisation angles derived at 4.8~GHz ($\chi=61\fdg7$) and
8.4~GHz ($71\fdg3$) from our calibration procedure are very close to the
NRAO measurements made 17 days before ($\chi=58\fdg3$ at 4.8~GHz and
$\chi=68\fdg5$ at 8.4~GHz) and 30 days after ($\chi=58\fdg7$ at
4.8~GHz and $\chi=70\fdg9$ at 8.4~GHz) our observing epoch.   
This confirms that our calibrations of the electric vector
polarisation angles are accurate to within about $5\degr$.
 
\vspace{-0.06cm}
To derive {\it RM} information, images were obtained at
the different frequencies by using the {\it uv}-range between
the shortest baseline at the highest frequency and the longest baseline at
the lowest frequency, and convolving to the lowest resolution. 
Then, a data cube in $\lambda^2$ at four (the maximum number of
  frequencies accepted by the AIPS task RM) observed frequencies
(4619, 4854, 5094, and 8421~MHz) was obtained using the AIPS
task MCUBE.
The AIPS task RM was used to produce a weighted fit of the observed
$\chi$ values to a $\lambda^2$ dependence. Finally, the intrinsic
magnetic-field directions were obtained using the derived {\it RM}
distribution to de-rotate the observed $\chi$ values for the VLBI
polarisation distribution.

\subsubsection{Archival data}
\label{subsubsec:literature}
We also reanalysed data from projects BC033 ~\citep[May 1995;][]{Nan00}
and BZ0023 ~\citep[August 2000;][]{Zhang04} at C- and X- bands,
respectively, to obtain a more accurate determination of {\it RM}.
The C-band observations by \citet{Nan00} contained 
4 independent frequencies, namely 4650,  
4845, 4853, and 5052~MHz, while the X-band observations by \citet{Zhang04}
were centred on 7904, 8238, 8562, and 8888~MHz. 
This allows us to determine the rotation measure of the source by
using 8 and 5 data points at C- and X-band, respectively.
 
We calibrated the data using the same strategy as
described in Sect.~\ref{subsec:myobservation}.
Images were made with a restoring beam of $2.0\times 1.5$~mas
at P.A. $-6\degr$ and the same {\it uv}-range sampling as for our data, 
to perform a pixel-by-pixel comparison. The resultant images look very
similar to those obtained from our own data. The total flux densities that
we obtained agree (within 3\%) at both C- and X- bands. 

\section{Results}
\label{sec:results}

\subsection{The C-band results}
\label{subsec:5ghz}

\begin{figure}[hb]
      \centering
      \resizebox{\hsize}{!}
      {\includegraphics{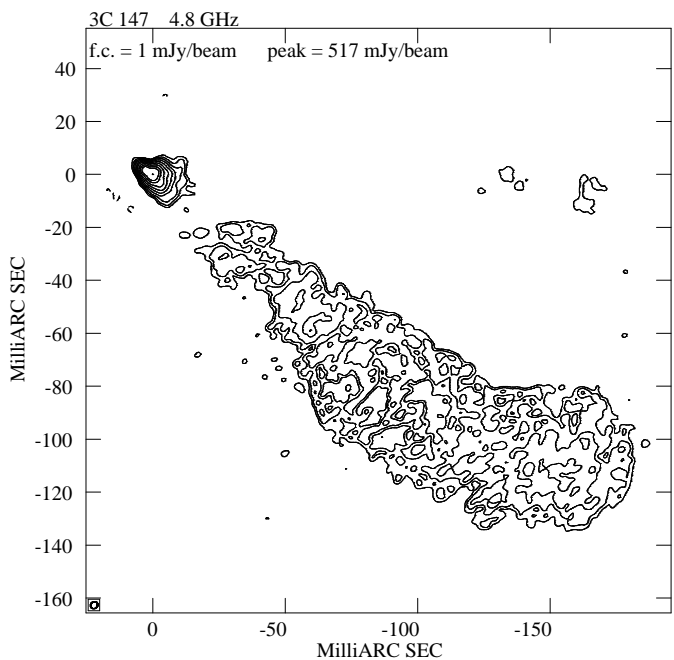}}
      \caption{The total intensity 4.8-GHz image of \object{3C\,147} using all
        four C-band IFs and a restoring beam of
      $2.8\times 2.5$~mas at $-40\degr$. Contour levels
      increase by a factor of 2 from 1~mJy/beam. 
	\label{fig:3c147big}}
    \end{figure}

The total-intensity image of \object{3C\,147} at 5.0~GHz obtained using all the
four C-band frequencies is
shown in Fig.~\ref{fig:3c147big}. This image was tapered to
improve the sensitivity to low brightness emission. The
off-source rms noise level achieved in this image is about
0.2~mJy/beam, just over a factor of two above the expected thermal noise.


We measured a total flux density of 4.70~Jy, which is $57\%$ of the
single-dish flux density \citep{Pauliny78} implying that a large 
fraction of the flux density was resolved.  
On the mas scale, \object{3C\,147} shows a one-sided structure with a complex 
compact region, and a component extending to the southwest, which is
visible out to a distance of $\approx 200$~mas from the compact
component. This soon broadens without losing its collimation, shows
several slight wiggles and, finally, bends to the west close to its end. 
    \begin{figure}[htbp]
      \centering
      \resizebox{\hsize}{!}
      {\includegraphics{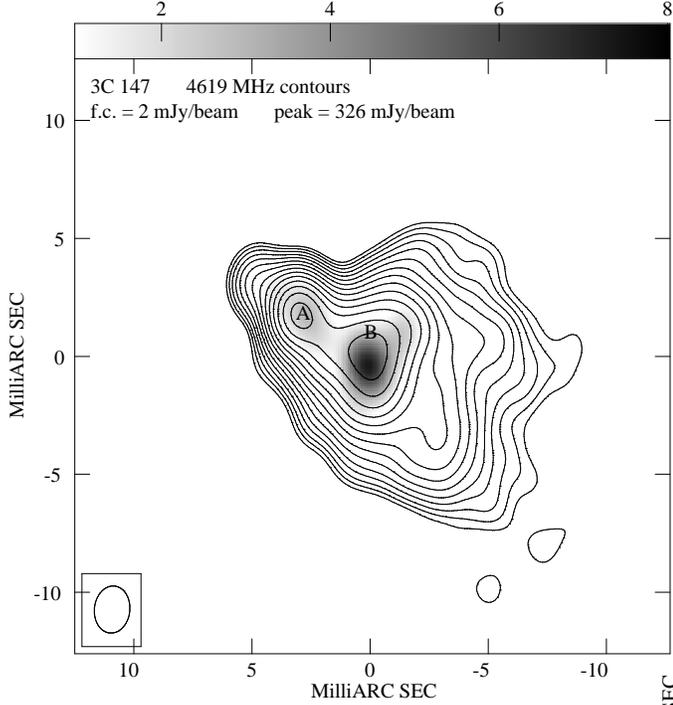}}
      \caption{The grey-scale image of C-band polarised intensity with contours
      of total intensity for IF\,1 (4619~MHz) superimposed. The
      angular resolution is $2.0\times 1.5$~mas at
      P.A. $-6\degr$. Contour levels increase by a factor of $\sqrt2$
      from 2~mJy/beam. 
	\label{fig:3c147polC}}
    \end{figure}
To compare the 5 and 8.4~GHz results, the images need to be of the
same resolution. Ideally, the {\it uv}-coverages should be identical. However,
this is impossible with VLBI arrays. Therefore, images of the central
component were made using the {\it uv}-range between the
shortest baseline at X-band and the longest baseline at C-band and
a restoring beam of $2.0\times 1.5$~mas at P.A. $-6\degr$. 

At C-band, we made images of the central component for each of the four
individual IFs.

\begin{table}[htbp]
\centering
 \caption{Source parameters.}
 \label{tab:fluxes}
\tabcolsep 1.4mm
 \begin{tabular}{cccccc}
 \hline
 \hline
Comp.     &  IF  &    S      & $S_p$      &  $\chi$     & \%pol    \\
          &  MHz &  mJy      &   mJy      &  $\degr$    &          \\
\hline
\hline 
  A       & 4619 & $335\pm10$& $3.3\pm0.3$&$-7  \pm5$  & $1.0 \pm 0.1$\\      
          & 4657 & $327\pm10$& $3.1\pm0.3$&$-1  \pm5$  & $0.9 \pm 0.1$\\
          & 4854 & $331\pm10$& $3.0\pm0.3$&$15  \pm5$  & $0.9 \pm 0.1$\\
          & 5094 & $319\pm10$& $2.9\pm0.3$&$29  \pm5$  & $0.9 \pm 0.1$\\
          & 8409 & $331\pm10$& $7.1\pm0.6$&$196 \pm5$  & $2.1 \pm 0.2$\\
          &      &           &            &            &              \\
  B       & 4619 & $596\pm18$& $8.9\pm0.4$& $-47 \pm5$ & $1.5 \pm 0.1$\\
          & 4657 & $579\pm17$&$10.5\pm0.4$& $-44 \pm5$ & $1.8 \pm 0.1$\\
          & 4854 & $587\pm18$&$12.1\pm0.4$& $-14 \pm5$ & $2.1 \pm 0.1$\\
          & 5094 & $565\pm17$&$10.6\pm0.4$& $+18 \pm5$ & $1.9 \pm 0.1$\\
          & 8409 & $491\pm15$&$31.7\pm1.0$&$+161 \pm5$ & $6.5 \pm 0.4$\\
\hline
\hline
\end{tabular}
\end{table}

Figure~\ref{fig:3c147polC} shows the structure of the central
region for IF\,1 (4619~MHz). The image reveals two compact components, A and
B, embedded in more diffuse emission.
The grey scale represents the distribution of polarised intensity. 
Polarised emission is detected from both components A and B,
the overall degree of C-band polarisation being about 1.0\% and 2.0\%
for  components A and B, respectively, (see Table~\ref{tab:fluxes}).
Flux densities were determined using the task IMEAN on 
the same region of the P and I images. In Table~\ref{tab:fluxes}, the total
flux densities, S, the polarised flux densities, S$_p$, the electric vector 
position angle, $\chi$ and the fractional polarisation, \% pol, are
listed for the four C-band IFs and for the X-band.

\subsection{The X-band results}
\label{subsec:8ghz}

\begin{figure}[htbp]
      \centering
      \resizebox{\hsize}{!}
      {\includegraphics{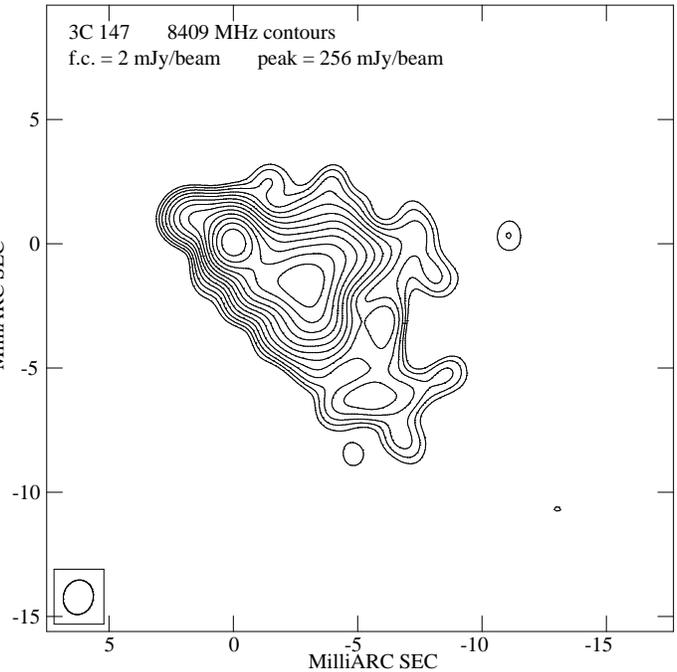}}
      \caption{Total intensity contours of \object{3C\,147} at 8.4\,GHz using
        all four X-band IFs and a restoring beam of $1.4\times 1.2$
        mas at P.A.$-14\degr$. Contour levels increase by a factor of $\sqrt2$ from 2~mJy/beam. 
	\label{fig:3c147-Xfull}}
    \end{figure}

The full resolution total-intensity image of the central
  component of \object{3C\,147} obtained using all the contiguous IFs at 8.4~GHz is given in
Fig.~\ref{fig:3c147-Xfull}. At this resolution the diffuse emission from
the outer part of the extended component is completely resolved.
The rms noise level in this image is 0.3~mJy/beam and we measured a
total flux density of about 1.6~Jy, $\approx 33\%$ of the
single-dish flux density \citep{Rickett06}.
  
The structure of the compact component changes little between the X-
and C-band. In Fig.~\ref{fig:3c147-Xfull}, the source shows
the two compact components A and B embedded in a region of diffuse 
emission as at C-band. However, this image has a more
complex morphology with two bulges of emission, BN and BS, for
component B and a slightly resolved feature, A$_0$, to the northeast of
component A.

\begin{figure}[htbp]
      \centering
      \resizebox{\hsize}{!}
      {\includegraphics{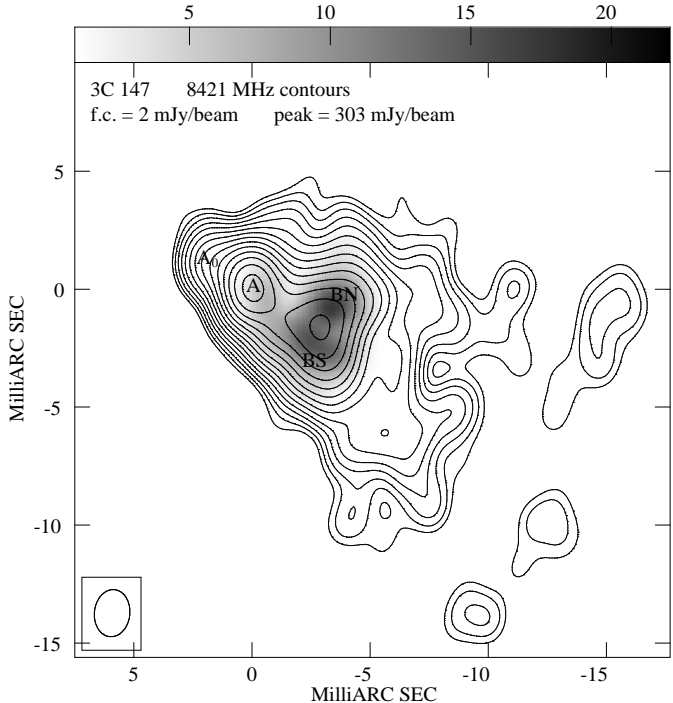}}
      \caption{Total intensity contours at 8421~MHz with a
      grey-scale image of polarised intensity superimposed.
      The angular resolution for both total and polarised intensities is as for Fig.~\ref{fig:3c147polC}.
      Contour levels increase by a factor of $\sqrt2$ from 2~mJy/beam. 
	\label{fig:3c147polX}}
    \end{figure}

The polarised flux density, whose distribution is represented in
Fig.~\ref{fig:3c147polX} in grey scale, reflects the complex structure
of the central region. Component A is polarised at a level of about 2\%. 
The peak of the X-band polarised intensity is not coincident with the
peak polarised intensity found at C-band. This is also seen  in other
sources that show evidence of {\it RM} gradients
(e.g., \object{3C\,119};~\citealt{Nan99}). 
For component A$_0$, we do not detect polarised emission.

Component B is resolved into two separated polarised regions 
associated with peaks of emission to the northwest (BN) and to the
south (BS). In these two regions, the fractional polarisation reaches
values of 6.6\% and  5.6\%, respectively. 
\vspace{-0.08cm}

\subsection{Comparison of C- and X-band results}
\label{subsec:comparison}

To compare C- and X-band images, they must be
aligned with each other. Absolute positional information is lost by the use of
self-calibration techniques, and the 5- and 8.4-GHz images were instead
aligned relative to each other by forcing the positions of the peak brightness
of component B to be coincident by means of the AIPS task LGEOM.

The spectral index values of the central components were obtained 
for a 5-frequency data cube using the AIPS task SPIXR.

\begin{figure}[htbp]
      \centering
      \resizebox{\hsize}{!}
      {\includegraphics{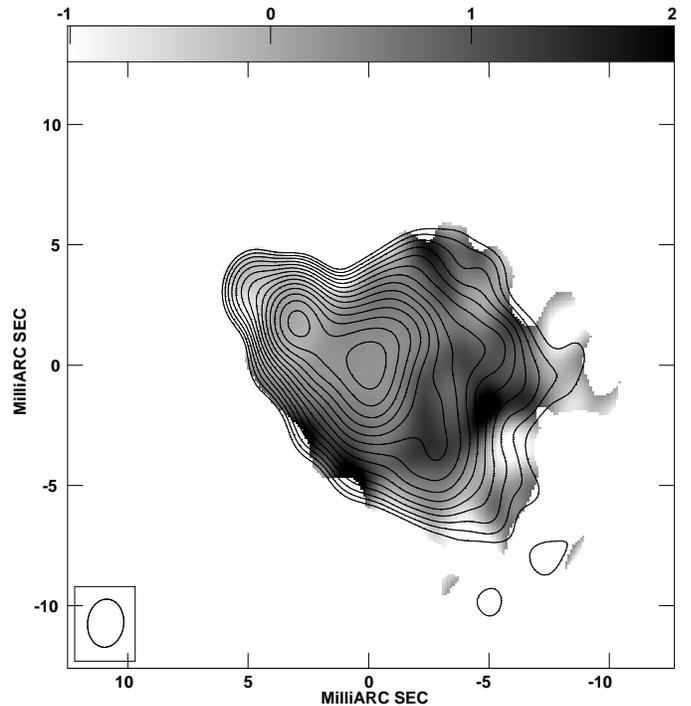}}
      \caption{A grey-scale image of the spectral-index distribution
      with the total-intensity contours for C-band IF\,1 (4619~MHz)
      superimposed. Contour levels increase by a factor of $\sqrt2$
      from 2~mJy/beam. 
	\label{fig:3c147spix}}
    \end{figure}

The spectral indices that we estimated are
$\alpha_{5}^{8.4}\approx 0.04$ for component
  A and $\alpha_{5}^{8.4}\approx 0.31$ for component B. The northeastern,
  slightly resolved component, A$_0$, has an inverted spectral index
  $\alpha_{5}^{8.4}\approx -0.4$. The overall spectral-index distribution
  is displayed in  Fig.~\ref{fig:3c147spix} in grey-scale.  

\subsection{Rotation measure structure and intrinsic magnetic field}
\label{subsec:rm-and-b}

    \begin{figure}[t]
      \centering
      \resizebox{\hsize}{!}
      {\includegraphics{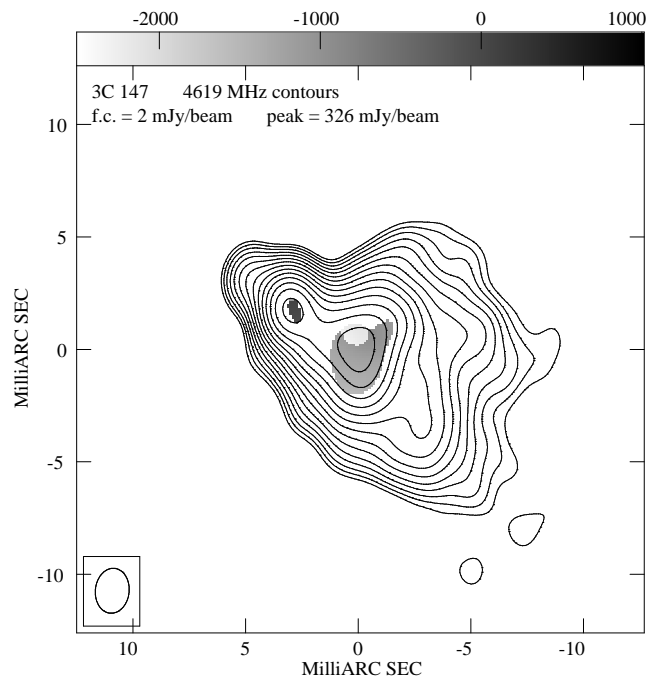}}
      \caption{Total-intensity C-band contours for IF\,1 (4619~MHz)
        with the {\it RM} distribution superimposed in grey scale.     
	\label{fig:3c147_rm}}
    \end{figure}

\begin{figure}[ht]
      \centering
      \resizebox{\hsize}{!}
      {\includegraphics{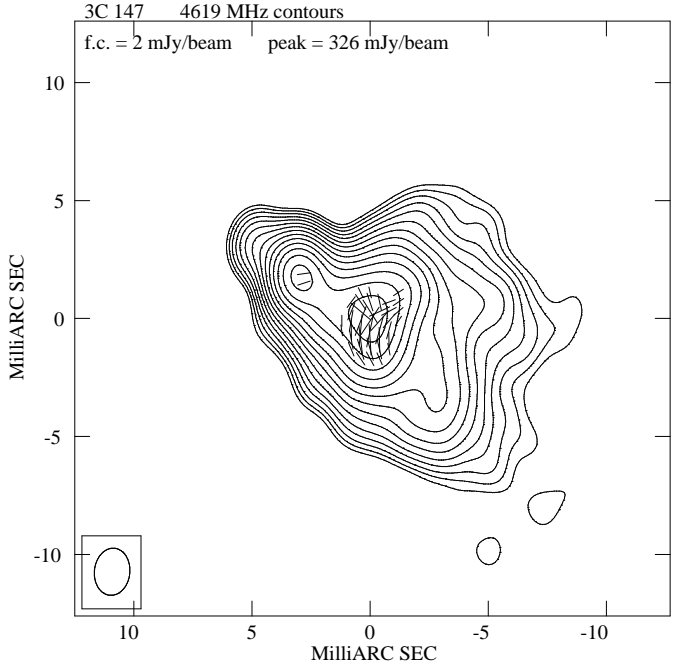}}
      \caption{Total-intensity C-band contours for IF\,1 (4619~MHz) with the
      intrinsic orientation of the projected {$\bf B$}-field superimposed.  
	\label{fig:3c147_bfield}}
    \end{figure}

Figure~\ref{fig:3c147_rm} shows in grey-scale the
{\it RM} distribution for \object{3C\,147} across four frequencies
overlaid on the total-intensity contour image.   
No redshift corrections were applied to the wavelengths, so the
{\it RM} in the rest frame of the source will be higher by a factor of
$(1+z)^2\approx 2.4$.
The {\it RM} distribution is not uniform but exhibits structure. The
highest {\it RM}s are associated with the northern polarised region of
component B.
The distribution of intrinsic magnetic-field orientation
is shown in Fig.~\ref{fig:3c147_bfield}. The magnetic-field structure
is complex and in the central component is perpendicular to the
major axis of the source. 

    \begin{figure}[htbp]
      \centering
      \resizebox{\hsize}{!}
      {\includegraphics{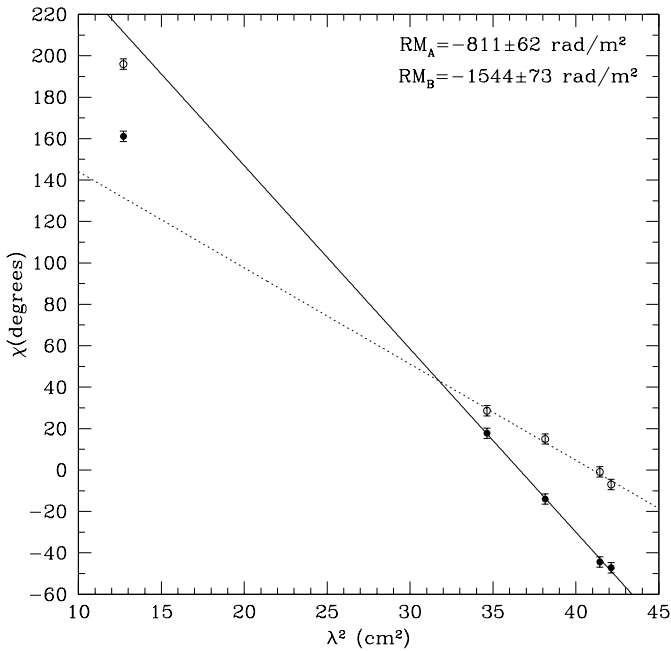}}
      \caption{A plot of the observed $\chi$ values for components A
      (open circles) and B (filled circles) as a function of $\lambda^2$ for
      the five available wavelengths.    
	\label{fig:rm_my}}
    \end{figure}

As mentioned in Sect.~\ref{sec:intro}, the behaviour of the
  electric vector polarisation angle, $\chi$, as a function of
$\lambda^2$ allows us to obtain information about the density distribution of 
the ionised ISM around the radio source. 
For a homogeneous (both in density and magnetic field) medium or 
a medium with inhomogeneities resolved by the observing beam, the
polarisation angle is strictly proportional to $\lambda^2$ at all
wavelengths. For an unresolved or partially resolved medium, the {\it RM}
changes from point to point across the source, different contributions
of polarised radiation are rotated differently, and the polarisation
angle can deviate from the  $\lambda^2$-linear law.

Figure~\ref{fig:rm_my} presents the polarisation angle of components
A(open circles) and B (filled circles) as a function of $\lambda^2$.  
Among the four C-band IFs, we found a good linear fit with
$RM_{\rm A}=-811\pm62$\,rad\,m$^{-2}$ and
$RM_{\rm B}=-1544\pm73$\,rad\,m$^{-2}$. However, the 8.4-GHz data
points are offset by
about $+64\degr$ and $-45\degr$, respectively. 

To investigate the {\it RM} structure of \object{3C\,147} and the
deviation of the electric vector polarisation angle from  the
$\lambda^2$-law further, we searched the VLBA data archive for 
available data to enable a direct comparison with our observations.

\subsection{Comparison of observations and archival data results}
\label{subsec:archiveresults}
The direct comparison between our images and those made from
archival data allows us to derive some interesting results.
\citet{Aller03} reported small systematic changes in flux density on a timescale of
several years. They suggested that the central region is responsible for
this intensity variability and also found variability in the
integrated polarised flux density. 

At C-band, we found that the total flux density we measured in our images is consistent
with the flux density measured in images
from project BC033, within the calibration uncertainties. 
However, between the two observing epochs
there are significant differences in the physical properties of the
resolved components that we summarise as follows:
\begin{itemize}

\item We imaged the difference in total-intensity between
the 1995 data from project BC033 \citep{Nan00}, and our data at the same
frequency, i.e., for IF\,3 of each data set (Fig.~\ref{fig:difference_C}).
The peak brightnesses of component A at the two epochs are not
coincident. Since the images were aligned by the position of component
 B, this indicates that the separation between components A and B has
 increased.  

    \begin{figure}[t]
      \centering
      \resizebox{\hsize}{!}
      {\includegraphics{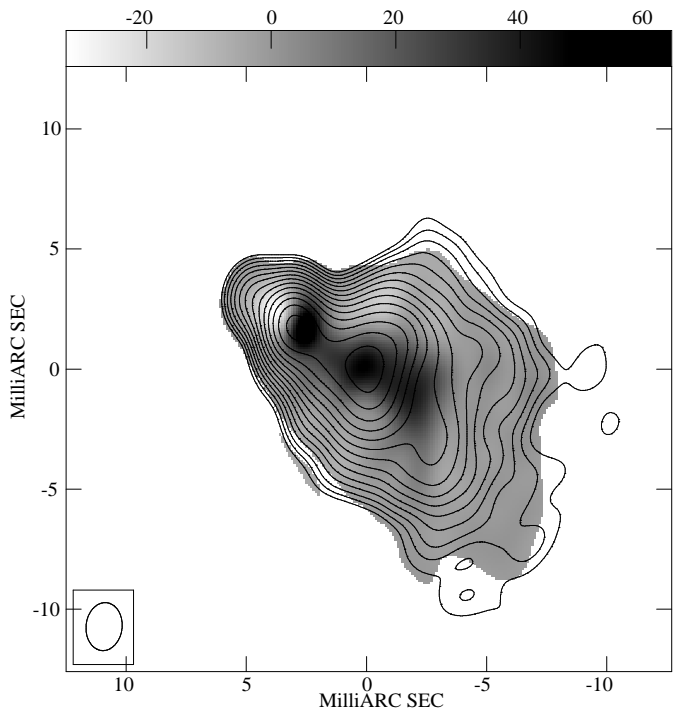}}
      \caption{A grey-scale image of the distribution of the C-band,
        total-intensity differences between 1995 and 2001,
      with contours of total intensity for IF 3 (4854~MHz)
      superimposed. Contour levels increase by a factor of $\sqrt2$. 
	\label{fig:difference_C}}
    \end{figure}

\item In \citet{Nan00}, there were indications that component A is not
polarised (with an upper limit to the fractional polarisation of about
0.5\%), and we confirm this result for images that we made from
their data. Based
on the lack of polarised emission, they identified component A with the
core. However, from our images of the epoch 2001 data, we measured a
polarised flux density associated with 
component A of about 3~mJy (see Table~\ref{tab:fluxes}), with a noise
level of 0.4~mJy/beam, i.e., a 
factor of eight above the rms, implying the presence of polarised
emission at the 1\% level. Therefore, we inferred a significant increase
in the C-band fractional polarisation associated with component A
between 1995 and our observing epoch of 2001.

\end {itemize}

\begin{figure}[htbp]
      \centering
      {\includegraphics[width=7.2cm]{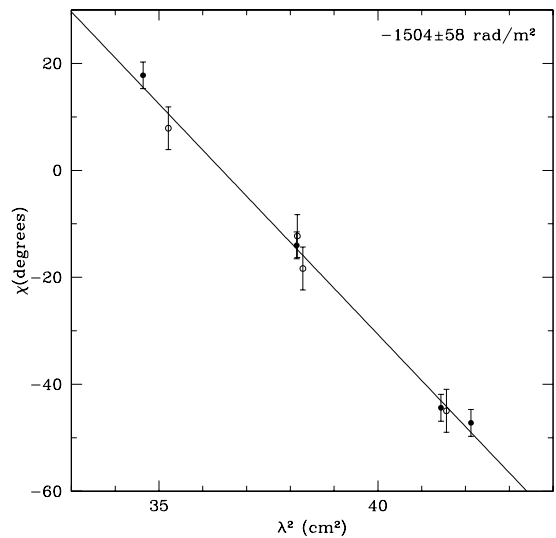}}
      \caption{A plot of the observed $\chi$ values for component B
      as a function of $\lambda^2$. Filled circles denote the four C-band
      IFs of our observations, while the open circles are the four
      C-band IFs from \citet{Nan00}.
	\label{fig:rm_new}}
    \end{figure}

Component B does not show any changes in its polarisation properties
between 1995 and 2001, and the electric vector angles derived for the
two data sets at C-band are in good agreement, giving a local {\it RM} of
$-1504\pm58$~rad\,m$^{-2}$. Figure~\ref{fig:rm_new} shows the electric
vector polarisation angle as a function of $\lambda^2$ at C-band for
component B as a whole by using the four IFs from project BC033 (open
circles) and the four IFs from our data (filled circles).

At X-band, we found no substantial differences between our data and
data from project BZ0023 \citep{Zhang04} for both components.  
Since these observations were made at similar epochs (May 2001 and Aug
2000), this gives us confidence in the reality of the C-band changes
detailed above.

    \begin{figure}[ht]
      \centering
      \resizebox{\hsize}{!}
      {\includegraphics{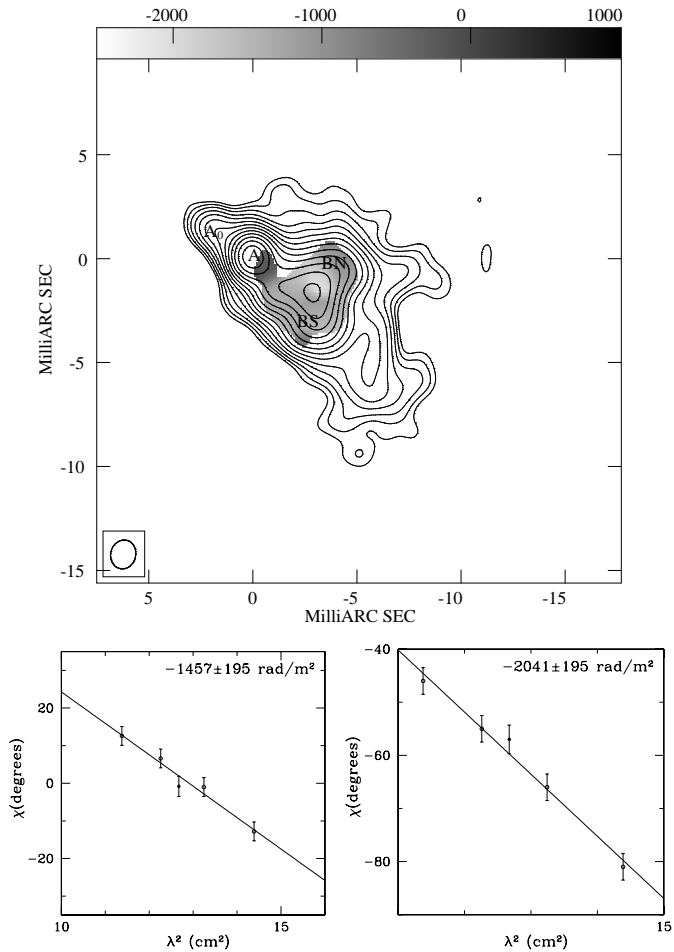}}
      \caption{Contours of X-band total intensity, with the {\it RM}
      distribution from project BZ0023 \citep{Zhang04} superimposed in
      grey scale. The two inserts show plots of $\chi$ versus
      $\lambda^2$ for the four IFs of the Zhang observation and our
      single X-band data point for the two polarised regions in
      component B, (the southern component to the left, and the
      northern component to the right).    
	\label{fig:3c147_Xrm}}
    \end{figure}

In Fig.~\ref{fig:3c147_Xrm}, we show the position angle of the electric
vectors as a function of $\lambda^2$ for the two polarised regions in component
B. The plots were obtained by using data from project BZ0023
(open circles) and
our single data point (filled circle). These are fitted rather well
with {\it RM}s of 
$-1457\pm195$~rad\,m$^{-2}$ and $-2041\pm195$~rad\,m$^{-2}$ for the
features to the  south and north respectively. 
Within the errors produced by noise and fitting to the
polarisation position angles, these {\it RM} values agree with the
single-dish measurements by \citet{Inoue95}. Therefore, the compact
region can be considered to be the major component responsible for the
high integrated {\it RM} of this radio source.

\subsection{Relativistic motion}
\label{subsec:motion}

\citet{Nan00} found indications of an increase in the
separation between components A and B. By comparing their data with
the results of~\citet{Alef90}, they found an apparent speed of $\approx
1.5\, c$.
To investigate this separation further, we
analysed the position of the components by fitting a Gaussian to
each component (in images obtained by using a selected range
of visibilities to sample the compact components only and a restoring
beam of $1\times1$~mas) and determining the distance between the peaks
of A and B. 
We found that the separation between the two components increased
by $0.6\pm0.2$ mas between May 1995 and May 2001, as measured from the 5-GHz
images for which the separation of epochs was the largest. This
corresponds to a relative motion between components A and B with an
apparent speed of expansion of $1.2\pm0.4\,c$. 


\section{Discussion}
\label {sec:discussion}

\subsection{Interpretation of the $\lambda^2$ linear law deviation}
\label{subsec:core}

The deviation from a $\lambda^2$ law observed in component B between C- and
  X-band (Fig~\ref{fig:rm_new}) can be explained in terms
of the contributions of two or more polarised components with
different $RM$s. Component B clearly shows evidence of
two compact polarised regions  with different {\it RM}s (see 
Figs.~\ref{fig:3c147polX} and~\ref{fig:3c147_Xrm}).
Therefore, at 8.4~GHz, the polarisation
angle that we are accounting for is an ``average value''. Neglecting other
effects, it equals the angle of the vector $\vec{\rm P}$ obtained by
 the vectorial sum of the two contributions.  
In such a situation, the polarisation angle is not expected to
follow a $\lambda^2$ law ~\citep{Rossetti08}.
An analogous interpretation could also explain the deviation from the
$\lambda^2$ law observed in component A, although this component was
not resolved into sub-components by the present observations.

\subsection{Possible explanations of the source structure}
\label{subsec:explanations}

\subsubsection{Scenario 1: the central region of \object{3C\,147} harbours the
  core and jet}

The standard interpretation of the radio structure of \object{3C\,147} is that
the central bright radio-emitting region contains the source nucleus.
This scenario is supported by the following evidence:
\begin{itemize}
\item The presence of two bright compact components, one with a flat
  spectrum (A) between 5 and 8~GHz, can be interpreted as a core
  plus a bright knot in the jet \citep{Nan00}. \citet{Ludke99} and
  \citet{Zhang04} detected an additional slightly resolved component
  to the
  northeast of component A, and, based on its weak polarisation and
  its position at one end of the source structure, they identified it
  to be
  the true core. If the core has lower polarisation than the knots, as
  found in core-dominated quasars \citep{Cawthorne93},
  milliarcsecond-resolution polarisation images should be very useful
  in identifying
  the core in \object{3C\,147}. However, although the northeast component
  A$_0$ is not completely resolved by the available observations,
  \citet{Zhang04} estimated its flux density to range from 32 to 26~mJy
  at 7904 and 8888~MHz respectively. The peak polarised signal is less
  than 1~mJy/beam, suggesting that the degree of polarisation is less
  than $\approx 3$\%.  
\item The triple structure found at subarcsecond resolution, which
  suggests the existence of a compact central core in-between two
  extended, asymmetric lobes, which are probably expanding within different
  environments \citep{Junor99a}.     
\end{itemize} 

   \begin{figure}[htbp]
      \centering
      \resizebox{\hsize}{!}
      {\includegraphics{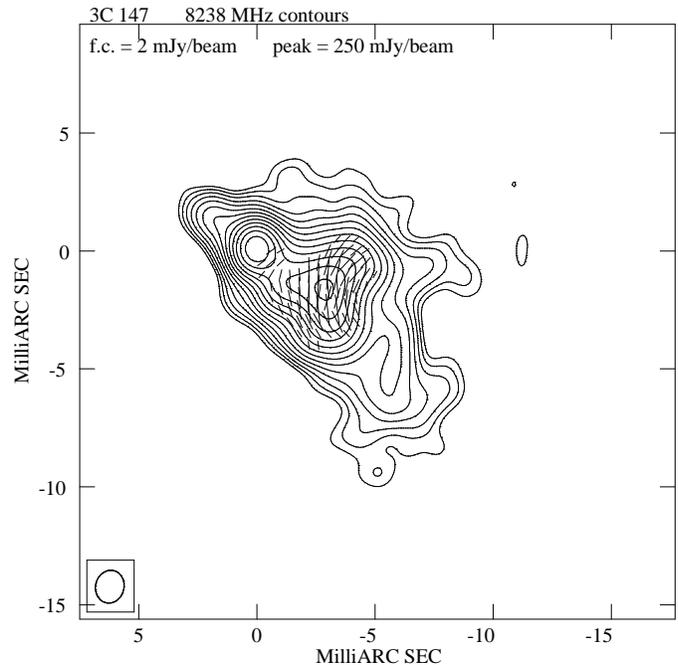}}
      \caption{Total intensity contours at 8238~MHz with the
      intrinsic orientation of the projected {$\bf B$}-field overlaid.  
	\label{fig:3c147_Xbfield}}
    \end{figure}

It is important to assess whether this scenario is consistent with
the magnetic field orientation along the axis of the jet
(Fig.~\ref{fig:3c147_Xbfield}). 

Although early work suggested that in quasars, the magnetic field is
often parallel to the jet axis \citep{Cawthorne93, Dallacasa95},
more recent surveys identified a more complex situation. 
For example, from polarimetric observations of 177 sources from the
Caltech-Jodrell Bank Flat-Spectrum Survey, \citet{Pollack03} found no
evidence of field direction alignment with the jet axis.  
Similar conclusions were reached by \citet{Lister05} based on
milliarcsecond-scale linear polarisation observations at 15~GHz of 133
jets associated with active galactic nuclei. 
Field lines on parsec scales can be complex in
radio jet$-$ambient medium interactions, as seen, for example, in the
quasar \object{B$1055+018$} \citep{Attridge99}.
Therefore, the magnetic field orientation in the compact region of
\object{3C\,147}, where there is possibly strong jet$-$medium interactions,
would not be unusual.

However, the identification of the source core still remains questionable.
Observation of far higher resolution were carried out by
\citet{Lister01} at 43~GHz for the central region of \object{3C\,147}. A bright,
compact source of size about 0.25~mas with a flux density of 177~mJy
and an upper limit to the degree of polarisation of 3.2\%, was
detected by these observations. However, this
component cannot be clearly identified with any of the features found
at lower frequencies within the source. It may correspond to either
of our components A or  A$_0$; its measured flux density is compatible
with the flat spectrum found for component A between 5.0 and 8.4~GHz. 
Although component A shows polarised emission, its value is consistent
with the measurements of quasar core polarisation, for which values can be
as high as 10\% , but the typical value is 
approximately a few per cent \citep{Pollack03, Lister05}.  

On the other hand, if A$_0$ is identified with the component seen at 43~GHz 
by \citet{Lister01}, it would have an inverted spectrum with a spectral index 
of approximately $-1$. This would be consistent with the most inverted spectrum
component, located at one end of the jet, being the core, as also seen
in  \object{3C\,48} \citep{Wilkinson91}. However, the spectrum should 
turn over so that the extrapolated flux density at 230~GHz does not exceed the
total flux density value of $280\pm28$~mJy determined by \citet{Steppe95}.

\subsubsection{Scenario 2: the compact region of \object{3C\,147} is a
  hot spot complex}

A second interpretation that we discuss for the radio structure of \object{3C\,147}
is that the brightest region imaged in Fig.~\ref{fig:3c147polC} is the
approaching hot spot of a powerful radio source, whose main axis is
quite close to the line of sight. 
The hot spot is marginally Doppler boosted, and its
structure, with multiple, individual knots, is similar to the multiple
hot spots found in several powerful double radio galaxies when studied
with appropriate resolution. 
What was previously defined as a ``jet''  is now interpreted as the backflow tail of the
approaching lobe.  
The other lobe lies about $1\farcs2$ to the north
\citep{Junor99a} and its hot spot is Doppler dimmed, since particle
acceleration takes place on its receeding side.
The strong asymmetry between the two lobes in both size and flux
density may  arise from differing environments, as suggested by 
\citet{Junor99a}. In that case, the main lobe appears brighter and
smaller because of stronger confinement of the radio-emitting
plasma. If the brightest lobe is confined by a far denser medium, we
would expect ionization of the ambient medium in the region where
interaction with the radio jet takes place. This 
would explain the high Faraday rotation that we observe.

The overall properties of \object{3C\,147} are similar to those found in
\object{3C\,286} \citep{Jiang96, Cotton97}, and possibily
  \object{3C\,138} \citep{Cotton03b}
(although both of these sources are highly polarised with low/absent {\it
    RM})
whose interpretation presents similar challenges.

Evidences that are consistent with this lobe-hot spot scenario  can be
summarised as follows: 

\begin{itemize}
\item The polarisation properties (magnetic field topology and fractional
polarisation) in the brightest region of the source are
consistent with a hot spot of multiple structures.

\item The extended structure to the southwest of the brightest
  region, which we propose to be a backflow tail, has a
  smooth brightness distribution; the structure is rather slim
  and is similar to the lobes in thin FR~II sources, such as \object{3C\,284}
  and \object{3C\,341}. On the small scale, it resembles the main ``jet'' of
  \object{3C\,138} \citep[e.g.,][]{Cotton03b}, which is not a jet, since the
  current interpretation of the source postulates 
that the easternmost and brightest structure is indeed a hot spot.

\item The size of the various components and the width-to-length ratio
of the brightest region can be more easily reconciled with a hot spot rather
than the initial part of a jet (i.e., the core), since this would require
substantial recollimation to maintain a small opening angle along the entire structure to the southeast.
We note that there are some examples of jets
exhibiting large opening angles and later recollimating, as seen in
\object{M\,87} \citep{Junor99b} and, to some extent, in
\object{B$1055+018$} \citep{Attridge99}, even if this was
found to occur on far smaller scales. 

\end{itemize}

However, a few aspects still require  
explanation because they do not fully agree with this ``hot spot''
interpretation.
\begin{itemize}
\item The core has yet to be found.

 Among the various possibilities, the
small-scale jet axis may be misaligned with the large-scale jet and, in
particular, be so close to the plane of the sky that the highly
relativistic jets are Doppler dimmed, or the radio
activity may have recently been switched off, as  proposed by
\citet{Kunert05} for the ``fader'' radio sources. The hot spots remain
active, while the plenishment of freshly created particles has ceased.

 Alternatively, a ``normal'' weak (steep-spectrum?) core appears to be 
embedded within the diffuse emission of the lobe+hot spot complex, and
cannot be clearly identified in present observations. Its contribution to
the total flux density of the radio source at cm wavelengths would be
marginal, given that the source is known to be quite stable in flux
density.

\item The northern lobe is difficult to explain if we consider a
model of a small-sized classical double structure, such as
\object{Cygnus\,A}, seen in
projection. However, there are powerful sources in which radio plumes
emerge from the classical double structure, and \object{3C\,147}
may resemble  either \object{3C\,249.1} or \object{3C\,351} seen
along the source major axis.

\item Superluminal motion has been observed
between the two compact components A and B. The apparent separation speed
of the knots in the compact component of \object{3C\,147} (see Sect.~\ref{subsec:motion}) is
consistent with the general model of AGN radio sources, in which the
jet is assumed to be initially highly relativistic. In case of a jet termination along the
line of sight, it might be possible to observe relative separation at the
apparent speed of light.

\end{itemize}

\subsection{Comparison with other polarised CSS quasars}

Only a minority of CSS quasars has been imaged so far with polarimetric VLBI
observations.
As an example, data have been published for only 7 of the 24 objects
in the list of CSSs from the 3C and PW 
catalogues \citep{Fanti90}.
All of these show a core-jet structure. Polarised emission is detected
along the jet, showing fractional polarisations of between 2\% and 20\%.
In contrast to
flat spectrum quasars the cores, when detected, are not polarised. 
CSSs for which  mas-scale {\it RM} distributions have been derived are
even rarer. We note here the cases of \object{OQ\,172}
\citep{Udomprasert97}, \object{3C\,216} \citep{Venturi99},
\object{$0548+165$} and  \object{$1524-136$} \citep{Mantovani02},
\object{3C\,43} and \object{3C\,454} \citep{Cotton03c}, and
\object{3C\,138} \citep{Cotton03b}. 
In those cases, rest-frame $RM > 10^3$\,rad\,m$^{-2}$, and even as
high as $2
\times 10^4$ are found.  
The fractional polarisation decreases moving from higher to lower frequencies.
Generally speaking, the {\it RM} is explained in terms of a foreground
Faraday screen, where a NLR sometimes contributes to the {\it RM}.
The intensity of the magnetic field is estimated to be a few tenths of a
$\mu$Gauss. 
Jets are often distorted and this cas also be interpreted in terms of jet-cloud
interactions or projection effects.

The images of \object{3C\,147} presented here show a relatively undisturbed
``jet'', which has only a few slight wiggles and no sharp turns and
distortions seen in many other CSS sources. The only bend visible in
our images is at the end of the component to southwest.
The observed high {\it RM} values are associated with the compact
region, while in many core-jet CSS quasars high integrated Faraday
rotation takes place in the bright region where the bend occurs.

\section{Summary and conclusions}
\label{sec:conclusions}

By analysing new polarimetric observations at C- and X-band, and 
re-analysing previous VLBA observations, we have been able to image the
milliarcsecond structure of the quasar \object{3C\,147}. 
We have identified a complex central region that is dominated by two bright
components, A and B, with a weak, compact component to the northeast,
A$_0$, and a further extended component to the southwest,
which is imaged out to a distance of 200\,mas.

Polarised emission has been detected for both of the compact
components, A and B, at C- and X-bands. There are indications of
polarised flux-density 
variability in component A on a timescale of several years. 
The polarised emission from component B divides into two independent 
regions in our observations. The two compact components also show very
high values of  {\it RM} in the rest frame of the source, these being
consistent with the rest-frame value reported for the entire main component by
\citet{Junor99a} of $RM =-3140\pm630$\,rad\,m$^{-2}$.

The intrinsic orientation of the magnetic field in the central region
of \object{3C\,147}  follows an arc-shape structure at the southern end of the
polarised region. The magnetic-field orientation is almost
perpendicular to the source major axis.
Similar results were obtained by \citet{Junor99a} at sub-arcsecond
resolution.
The separation between the two central
components of the source seems to be increasing with an apparent velocity of
$1.2\pm0.4\, c$. 

A straightforward interpretation of all the observational data for this
radio source is not yet possible.
We have discussed the possible location of the core and explored two possible
scenarios: a core-jet source, and a lobe-hot spot 
structure. Further polarisation images, and a more thorough
search for the source core, should help us to discriminate between these two
scenarios.

\begin{acknowledgements}

The VLBA is operated by the U.S. National Radio Astronomy Observatory
which is a facility of the National Science Foundation operated under
a cooperative agreement by Associated Universities, Inc.
We are very grateful to the referee for very helpful comments
and suggestions and for a careful reading of the manuscript of this paper.

\end{acknowledgements}

\listofobjects

\end{document}